\documentclass[aps,prd,twocolumn,amsmath,amssymb,nofootinbib,preprintnumbers]{revtex4}

\usepackage[pdftex]{graphicx}
\usepackage{epstopdf}
\usepackage{amsmath}
\usepackage{amssymb}
\usepackage{subfigure}
\usepackage{hyperref}
\usepackage{url}
\usepackage{xcolor}
\usepackage{color}
\usepackage{graphicx}
\usepackage{gensymb}
\usepackage{tabularx}

\begin{document} 

\title{\textbf{Trajectory of a flying plasma mirror traversing a target with density gradient}}

\author{Pisin Chen$^{1,2}$\footnote{pisinchen@phys.ntu.edu.tw} and
   Gerard Mourou$^{3}$\footnote{gerard.mourou@polytechnique.edu},
      }

 \affiliation{%
   ~\\
$^{1}$Leung Center for Cosmology and Particle Astrophysics \& Department of Physics and Graduate Institute of Astrophysics, National Taiwan University, Taipei 10617, Taiwan\\
$^{2}$Kavli Institute for Particle Astrophysics and Cosmology, SLAC National Accelerator Laboratory, Stanford University, CA 94305, U.S.A.\\
$^{3}$IZEST, Ecole Polytechnique, 91128 Palaiseau Cedex, France
}%

\begin{abstract}
It has been proposed that laser-induced relativistic plasma mirror can accelerate if the plasma has a properly tailored density profile. Such accelerating plasma mirrors can serve as analog black holes to investigate Hawking evaporation and the associated information loss paradox. Here we reexamine the underlying dynamics of mirror motion in a graded-density plasma to provide an explicit trajectory as a function of the plasma density and its gradient. Specifically, a decreasing plasma density profile (down-ramp) along the direction of laser propagation would in general accelerate the mirror. In particular, a constant-plus-exponential density profile would generate the Davies-Fulling trajectory with a well-defined analog Hawking temperature, which is sensitive to the plasma density gradient but not to the density itself. We show that without invoking nano-fabricated thin-films, a much lower density gas target at, for example, $\sim 1\times 10^{17}{\rm cm}^{-3}$, would be able to induce an analog Hawking temperature, $k_{_B}T_{_H}\sim 6.6 \times 10^{-2}{\rm eV}$, in the far-infrared region. We hope that this would help to better realize the experiment proposed by Chen and Mourou.

\end{abstract}

\maketitle

\newpage
 

\section{Introduction}

The black hole Hawking evaporation \cite{Hawking:1975} has triggered the wonder whether it would result in the loss of information \cite{Hawking:1978}. If yes, it would signalize a severe conflict between general relativity and quantum field theory; but if no, then how would the information be preserved \cite{Page:1993, Susskind:1993, AMPS:2013, Hotta:2015, Chen:2015, Hawking:2017, Hotta:2018, Chiang:2020}? This information loss paradox has been under debate for more than 40 years, which are essentially theoretical, without a firm conclusion. It is, however, generally agreed that the final verdict relies heavily on our knowledge about the end-stage of Hawking evaporation (For an overview, see, for example, \cite{Chen:2015} and references therein.). It can be easily verified that the Hawking lifetime of a stellar-size black hole is much, much longer than the age of the universe. Therefore the resolution to this paradox through direct astrophysical observations seems hopeless. Giving the fundamental importance of this issue, and that physics is essentially an experimental science, the idea of creating analog black hole in the laboratory for such investigation has become attractive. Unruh first proposed to investigate the Hawking evaporation through what he called the "dump hole" \cite{Unruh:1981}. More recently Steinhauer and co-workers have investigated a Bose-Einstein condensate (BEC) system as an analog black hole \cite{Steinhauer:2018}. 

The idea of a relativistic flying mirror as an analog black hole has been theoretically studied ever since Hawking's seminal discovery in 1975. It was Wilczek who first pointed out \cite{Wilczek:1993} that a flying mirror can further serve to investigate the information loss paradox. It, however, remains a gedanken experiment over the years until recently. In 2017, it was proposed \cite{Chen:2017} that such an experiment can indeed be realized through an accelerating plasma mirror induced by an intense laser that impinges a plasma with a graded density. The feasibility of this proposed scheme relies heavily on a more precise description of the mirror acceleration as a function of the local plasma density and gradient. 

It is well-known that in the plasma wakefield acceleration (PWFA) \cite{Chen:1985} and the laser wakefield acceleration (LWFA) \cite{Tajima:1979} mechanisms, the phase velocity of the plasma wakefield $v_{ph}$ equals the group velocity of the driver, $v_g$, be it a charged particle beam (in PWFA) or a short laser pulse (in LWFA), i.e., $v_{ph}=v_g$. This {\it wakefield principle} is true, however, only for a uniform-density plasma. In a more general situation where the plasma ambient density is nonuniform, the interplay between the laser group velocity and 
the wakefield phase velocity becomes more intricate and $v_{ph}$ and $v_g$ are not necessary equal. A generalized treatment of the plasma mirror dynamics should therefore break away from the conventional wakefield principle for uniform plasmas. 

In addition, in Ref.\cite{Chen:2017} the acceleration of plasma mirrors was derived based on the eikonal equation \cite{Timur:2017}, which is the foundation of geometric optics and corresponds to the zero-wavelength limit. The validity of the eikonal approximation to treat the acceleration in a nonuniform plasma lies in the assumption that the density gradient is minute. To apply to more general situations, a different, less restrictive formulation is desirable. 

Another motivation of this work is to search for a new and simplified operation regime for the proposed flying plasma mirror as analog black hole experiment \cite{Chen:2017}. In the original Chen-Mourou proposal, a two-stage laser-plasma interaction was invoked, where the second stage assumes a nano-fabricated thin-film target with graded density to induce the analog Hawking radiation. One practical concern is that such nano-target would generate excessive backgrounds that compete against the rare Hawking signals. The preparation of the high-intensity x-ray pulse is another technical challenge. Such high intensity, high frequency x-ray pulse is necessary for penetrating the high density nano-target and inducing the flying plasma mirror inside. One convenient way to prepare such x-ray pulse is to reflect an optical laser pulse against a relativistic plasma mirror \cite{Timur:2017}, which, however, has a rather low reflectivity. With these considerations in mind, it is highly desirable to explore the possibility of a single-stage, gaseous plasma target that could deliver the proposed analog black hole experiment.

Motivated by these considerations, in this paper we re-derive the acceleration of a plasma mirror, which in essence is a plasma wakefield, in the strongly nonlinear perturbation regime, induced by an ultra-short laser pulse based on the generalized wakefield principle. In reality, the shortest possible laser pulse length is one laser wavelength, i.e, a state-of-the-art single cycle laser pulse \cite{Mourou}. Furthermore, the laser wavelength must be much shorter than the plasma wavelength, or $\omega_p^2/\omega_0^2\ll 1$, such that the laser pulse length can be reasonably treated as a delta function. The first ion bubble size for a laser pulse with multiple cycles will necessarily be modified due to the intricate laser-plasma interaction, which we try to avoid.

The driving laser pulse, through dumping its energy into the plasma to create the wakefields, would suffer a frequency redshift and therefore would gradually slow down, or decelerate. This, however, is a slow process. In the case of a uniform plasma, such a slow-down will result in the slow-down of the plasma wakefield or mirror. The situation is different when the laser traverses a nonuniform plasma. 

There are several factors that determine whether a relativistic flying plasma mirror would accelerate or decelerate when traversing a plasma with a density gradient. In the case of an increasing/decreasing  density along the direction of the laser propagation, two effects would happen. One, in addition to the red-shift mentioned above, the driving laser will face an additional slow-down/speed-up mechanism by entering a denser/less-dense plasma region. Namely, as the local refractive index for the same laser frequency becomes smaller/larger, its group velocity decreases/increases. Two, upon entering a denser/less-dense region, the ambient plasma wavelength becomes shorter/longer and therefore the mirror, i.e., the wakefield, behind the driving pulse catches up/falls behind. Such dichotomy surely breaks the conventional wisdom of $v_{ph}=v_g$, which stems from the uniform plasma situations. Whether the net effect results in the acceleration or deceleration of the plasma wakefield or mirror depends on the interplay between the laser pulse and the local plasma conditions. 

Note that both the driving laser and the plasma mirror move with a velocity already very close to the speed of light. Since by construction the mirror accelerates, its velocity may at some point exceed the speed of light. No physical principle is violated, of course, since the plasma mirror moves with a phase velocity. With the motivation to mimic the black hole evaporation, in this paper we consider truncating the mirror motion before it turns superluminal, so as to release Hawking radiation's partner modes properly \cite{Wilczek:1993, Hotta:2015, Chen:2017}.\footnote{It is actually fine that the plasma mirror enters the superluminal regime before it stops. The partner modes would still be trapped until the mirror eventually stops. This superluminal regime may provide extra interesting black hole physics.}

This paper is structured as follows. In Sec.~\ref{sec:rad}, we analyze the equation of motion of a plasma mirror traversing a nonuniform plasma. In Sec.~\ref{sec:radgrav}, we investigate the variation of the ion bubble size. In Sec.~\ref{sec:mirroraccel} we derive the plasma velocity and trajectory as a function of the plasma density gradient. In Sec.~\ref{sec:mirroraccel} we derive the mirror acceleration and explore the parameter regimes where the mirror would accelerate. We then apply our results to the physics of analog black holes in Sec.~\ref{sec:analogBH}. In Sec.~\ref{sec:max}, we discuss the Hawking temperature induced by such accelerating mirror.
Finally, in Sec.~\ref{sec:dis}, we briefly comment on some experimental considerations for such an analog black hole experiment.

\section{\label{sec:rad}Competition between two parties}

In the nonlinear (blowout) regime of plasma perturbations, the back-end of the ``ion bubble" with the length  $x_{_B}$ that trails behind the driving pulse is where the plasma mirror locates. 
So the position of the plasma mirror is
\begin{eqnarray}
x_{_M}=x_{_L}-x_{_B},
\end{eqnarray}
where $x_M$ and $x_L$ are the position of the mirror and the laser, respectively, in the lab frame. Let $\dot{x}\equiv dx/dt$ and $\ddot{x}\equiv d^2x/dt^2$. Then the velocity and the acceleration of the mirror are
 \begin{eqnarray} 
\dot{x}_{_M}=\dot{x}_{_L}-\dot{x}_{_B},\\
\ddot{x}_{_M}=\ddot{x}_{_L}-\ddot{x}_{_B},
\end{eqnarray}
respectively. $\dot{x}_{_L}$ is what we usually identify as the laser group velocity, $v_g\equiv \eta c$, where the index of refraction $\eta$ of a plasma in the nonlinear perturbation regime is 
\begin{eqnarray}
\eta=\sqrt{1-\frac{\omega_p^2}{\omega_0^2}\frac{1}{1+\phi}},
\end{eqnarray}      
where $\phi$ is the normalized dimensionless scalar potential of the laser-plasma system. 

\section{\label{sec:rad}Speed Change of Driving Laser}

The rate of change of the laser group velocity $\ddot{x}_{_L}=\dot{v}_g$ is directly related to that of the index of refraction, which in general should depend on the variations of $\omega_p$, $\omega_0,$ and $\phi$:
\begin{eqnarray}
\dot{\eta}(x)=\frac{1}{\eta}\frac{\omega_p^2}{\omega_0^2}\Big\{ \Big( \frac{\dot{\omega}_0}{\omega_0}-\frac{\dot{\omega}_p}{\omega_p}\Big)\frac{1}{1+\phi}+\frac{\dot{\phi}}{2(1+\phi)^2}\Big\}.
\end{eqnarray}
Modeling the laser pulse envelop with a sine function, it can be shown \cite{Sprangle:1990,Esarey:1990} that 
\begin{eqnarray}
\phi\simeq \frac{a_{L }^2 k_{p}^2}{8} \Big\{\chi^2-2\Big(\frac{L}{2\pi}\Big)^2 \Big[1-\cos(2\pi\chi/L)\Big] \Big\},
\end{eqnarray} 
where $a_{L}$ is the (Lorentz invariant) laser strength parameter, $L\sim \lambda_0\ll \lambda_p$ is the laser pulse length, and $0\leq \chi \leq L$ is the comoving coordinate relative to the head of the laser envelope. We can see that $\phi\ll 1$ within the envelope of a very short laser pulse. Furthermore, the change of the potential within the laser pulse, $\Delta\phi/\Delta\chi\ll \phi/L$, is also minute. We are therefore safe to ignore the contributions of $\phi$ and $\dot{\phi}$ to $\eta$ when dealing with the acceleration or deceleration of the laser pulse, i.e.,
\begin{eqnarray}
\eta(x)\simeq \sqrt{1-\frac{\omega_p^2}{\omega_0^2}}.
\end{eqnarray} 
We emphasize that this approximation applies only to the refractive index within the laser envelop even though the laser is ultra-intense. This is a fortuitous situation due to the fact that the laser pulse is ultra-short. 

Let us examine the contributions to the change of the refractive index separately below. 

\noindent
1. {\it Change of $\eta(x)$ due to local density gradient}

Holding the laser frequency fixed, the acceleration (or deceleration) of the laser group velocity due to the variation of plasma density is
\begin{eqnarray}
\dot{v}_{g1}=c\dot{\eta}|_{\omega_0}\simeq -\frac{c}{\eta}\frac{\omega_p^2}{\omega_0^2}\Big(\frac{\dot{\omega}_p}{\omega_p}\Big).
\end{eqnarray}
We see that for an increasing plasma frequency in time, which is directly related to an increasing plasma density along the laser propagation direction, the laser pulse tends to decelerate.

\noindent
2. {\it Laser frequency redshift due to wakefield excitation}

The driving laser frequency redshift was well described in the concept of {\it photon accelerator} by Wilks et al. \cite{Wilks1989}. In typical laser-plasma interactions the laser photon number is roughly conserved. As a consequence the conservation of energy demands that, by exciting the plasma wakefields, the energy loss of the laser is manifested through a frequency redshift of all photons. In the limit where the laser frequency $\omega_0$ is much larger than that of the plasma, i.e., $\omega_0\gg \omega_p$, the laser frequency redshift per unit distance of propagation is  \cite{Wilks1989}
\begin{eqnarray}
\frac{\delta\omega_0}{\delta x}=-\frac{\omega_p^2k_p}{2\omega_0}\frac{\delta n_p}{n_p}.
\end{eqnarray}
Since 
\begin{eqnarray}
\frac{\delta n_p}{n_p}\simeq \frac{e
E}{m\omega_p c}=a_{_L}\frac{\omega_0}{\omega_p},
\end{eqnarray}
where $\omega_p\equiv \sqrt{4\pi r_en_p}$, we have, in the nonlinear laser-plasma interaction regime (i.e., $a_{_L}\geq 1$), 
\begin{eqnarray}
\frac{\delta\omega_0}{\delta x}\simeq -\frac{a_{_L}}{2}\omega_p k_p.
\end{eqnarray}
Holding $\omega_p$ constant,  that is, for a uniform plasma, the rate of change of the laser group velocity is
\begin{eqnarray}
\dot{v}_{g2}=c\dot{\eta}|_{\omega_p}\simeq \frac{c}{\eta}\frac{\omega_p^2}{\omega_0^2}\Big(\frac{\dot{\omega}_0}{\omega_0}\Big).
\end{eqnarray}
With $\dot{\omega}_0=c\eta \partial \omega_0/\partial x$, and identifying $\partial \omega_0/\partial x$ with $\delta \omega_0/\delta x$, we find 
\begin{eqnarray}
\dot{v}_{g2}\simeq -c\frac{a_{_L}}{2}k_p\frac{\omega_p^3}{\omega_0^3}.
\end{eqnarray}
Once again, this frequency redshift effect also slows down the laser. 

The total rate of change of the laser group velocity is the sum due to these two effects. Putting these two contributions together, the net deceleration of the laser pulse is 
\begin{eqnarray}
\ddot{x}_{_L}=\dot{v}_g=\dot{v}_{g1}+\dot{v}_{g2}= -\frac{c}{\eta}\frac{\omega_p^2}{\omega_0^2}\Big[\frac{\dot{\omega}_p}{\omega_p}+\frac{\dot{\omega}_0}{\omega_0}\Big]\\ \nonumber
\simeq -\frac{c}{\eta}\frac{\omega_p^2}{\omega_0^2}\Big[\frac{\dot{\omega}_p}{\omega_p}+\eta\frac{a_{_L}}{2}k_p\frac{\omega_p}{\omega_0}\Big].
\end{eqnarray}
Since $\omega_p^2\ll \omega_0^2$, we will save only terms up to $\omega_p^2/\omega_0^2$ in the rest of the discussion. Thus the deceleration due to the frequency redshift is relatively unimportant and we shall ignore the second term and regard $\omega_0$ is constant. By the same token, $\eta$ can be replaced by 1 in this expression.

\section{\label{sec:radgrav}Variation of Ion Bubble Size}

In this section we turn our attention to the variation of the ion bubble, that is, the change of $x_{_B}$ as the plasma density varies. The plasma wavelength is a subtle issue even in the case of uniform plasmas. In the absence of the driving laser, i.e., the source, the plasma perturbations, and therefore the wavelength, in all regimes (linear and nonlinear) is well described in the Akhiezer-Polovin theory \cite{Akhiezer:1956}. In the presence of the driving laser, however, the size of the first ion bubble immediately behind the laser is in general different from the Akhiezer-Polovin plasma wavelength due to the interplay between the laser pulse and the plasma. 

Laser-plasma interaction in the nonlinear regime has been well studied \cite{Sprangle:1990, Esarey:1990}. Relevant to our case of ultra-short, single-cycle laser pulse, it was shown that if the condition 
\begin{eqnarray}
\frac{\pi}{2}a_{_L}\frac{L}{\lambda_p}\sim \frac{\pi}{2}a_{_L}\frac{\omega_p}{\omega_0}\ll 1 ,
\end{eqnarray}
is satisfied, then the plasma wavelength excited by this short pulse follows the conventional expression in the linear regime, that is, 
\begin{eqnarray}
\lambda_p=\frac{2\pi c}{\omega_p},  \quad\quad {\rm even ~if~} a_{_L}\gg 1.
\end{eqnarray}
This fact can be intuitively appreciated by recognizing that when the laser pulse is ultra-short, the scalar potential within the laser pulse, $\phi$, is much less than unity (c.f. Eq.(6)). As a result, the ``transverse" Lorentz factor $\gamma_\perp\sim 1$. indicating that the plasma perturbations remain in the linear regime. It is ironic, but true, that in the ultra-short limit of the driving laser pulse the plasma wavelength in the nonlinear regime reduces to that in the linear regime, which significantly simplifies our subsequent derivations.

Furthermore, it can be mathematically proven \cite{Liu:2020} that the size (length) of the first ion bubble equals to 3/4 of the plasma wavelength (either linear or nonlinear) for a delta-function laser pulse. In the following we will assume such a limit in our derivations. That is,
\begin{eqnarray}
x_{_B}=\frac{3}{4}\lambda_p(x)=\frac{3\pi c}{2\omega_p(x)}.
\end{eqnarray}

The variation of the ion bubble size is therefore 
\begin{eqnarray}
\dot{x}_{_B}=-\frac{3\pi c}{2}\frac{\dot{\omega}_p}{\omega_p^2}.
\end{eqnarray}
We see that  $\dot{x}_{_B}$ is negative definite in the situation where $\dot{\omega}_p>0$, and is positive if $\dot{\omega}_p<0$. That is, the plasma wavelength, as well as the longitudinal bubble size, shrinks or grows when the laser traverses a plasma with an increasing or decreasing density, respectively.


\section{\label{sec:mirroraccel}Plasma Mirror Velocity and Trajectory}

In the concept of a moving mirror as an analog black hole \cite{Fulling-Davies:1976, Fulling-Davies:1977, Birrell:1983}, the mirror was assumed to be a real object and therefore should move with a velocity that is less than the speed of light. In the Chen-Mourou proposal, however, the mirror is not real but a phase wave, which can in principle have a speed faster than that of light. To qualify the plasma mirror as an analog black hole, we should ensure two conditions. First, the phase velocity of the plasma mirror is less than the speed of light, that is, $\dot{x}_{_M}/c < 1$. Second, it accelerates. 
 
Inserting $v_g$ and $\dot{x}_{_B}$ into Eq.(2), i.e., $\dot{x}_{_M}=\dot{x}_{_L}-\dot{x}_{_B}$, we obtain the velocity of the mirror:
\begin{eqnarray}
\frac{\dot{x}_{_M}}{c}=1-\frac{1}{2}\frac{\omega_p^2}{\omega_0^2}+\frac{3\pi}{2}\frac{\dot{\omega}_p}{\omega_p^2}.
 \end{eqnarray}
In our conception, the time derivatives of the plasma frequency are induced through the spacial variation of the plasma density via the relation $\omega_p(x)=c\sqrt{4\pi r_e n_p(x)}$. Since $d/dt=\partial/\partial t +(\partial x/\partial t)\partial/\partial x$ and $\omega_p$ does not depend on time explicitly, we find that $\dot{\omega}_p=v_{ph}\partial\omega_p/\partial x=\dot{x}_{_M}\partial\omega_p/\partial x\equiv \dot{x}_{_M}\omega_p'$. Substituting it into Eq.(19) and reshuffling terms, we find
\begin{eqnarray}
\frac{\dot{x}_{_M}}{c}=
\frac{1-(1/2)\omega_p^2/\omega_0^2}
{1-(3\pi/2)c\omega_p'/\omega_p^2}.
\end{eqnarray}

For the case of a decreasing density profile, a ``down ramp", i.e., $\omega_p'<0$, the sub-luminosity requirement is automatically satisfied. On the other hand for an increasing density profile, an ``up ramp", i.e., $\omega_p'>0$, the following constraint must be satisfied:
\begin{eqnarray}
\frac{\omega_p^{2}}{\omega_0^{2}} >
3\pi\frac{c\omega_p'}{\omega_p^2}.
\end{eqnarray} 
In the following we will consider only the down ramp case, where the the plasma density and its gradient are {\it not} constrained by the above condition.

The standard method to derive the response function of a particle detector in response to the perturbations induced by a flying mirror is based on the mirror's trajectory \cite{Birrell:1983}. With that purpose in mind, here we separate the $x$ and $t$ variables in Eq.(19) and integrate them separately, that is,
\begin{eqnarray}
\int_{t_0}^t \bar{c}dt=\int_{x_0}^x dx \Big(1-\frac{3\pi c}{2}\frac{\omega'}{\omega_p^2}\Big),
\end{eqnarray}
where $\bar{c}=(1-\omega_p^2/\omega_0^2)c=\eta c$ is the speed of light in the plasma medium. We find
\begin{eqnarray}
x(t)=x_0+\bar{c}(t-t_0)-\frac{3\pi c}{2}\Big[\frac{1}{\omega_p(x)}-\frac{1}{\omega_p(x_0)}\Big].
\end{eqnarray}
In Sec.VIII we will use this formula to estimate the Hawking temperature of a flying plasma mirror. 

\section{\label{sec:mirroraccel}Plasma Mirror Acceleration}

We now examine the acceleration of the plasma mirror. Again, since $\omega_p$ does not depend on time explicitly, we have 
\begin{eqnarray}
\dot{\omega}_p&=&\dot{x}_{_M}\omega_p', \cr
\ddot{\omega}_p&=&\ddot{x}_{_M}\omega_p'+\dot{x}_{_M}^2\omega_p'', \cr
{\dot{\omega}}'_p&=&\dot{x}_{_M}'\omega_p' +\dot{x}_{_M}\omega_p''.
\end{eqnarray}
Replacing $\dot{\omega}_p$ with $\dot{x}_{_M}\omega_p'$ and inserting $\dot{x}_{_M}$ from Eq.(20), we have,  
\begin{eqnarray}
\frac{\ddot{x}_{_M}}{c}&=&\frac{1-(1/2)\omega_p^2/\omega_0^2}{[1-(3\pi/2) c\omega_p'/\omega_p^2]^3}\times \\\nonumber
&&\Big\{-\frac{\omega_p^2}{\omega_0^2}\frac{\omega_p'}{\omega_p}\Big[1-\frac{3\pi c}{2}\frac{\omega_p'}{\omega_p^2}\Big] 
+\frac{3\pi c}{2}\Big[\frac{\omega_p''}{\omega_p^2}-2\frac{\omega_p'^2}{\omega_p^3}\Big]\Big\}.
\end{eqnarray}
The second and the fourth terms have the same dependence in $\omega_p'^2/\omega_p^3$, but the former is suppressed by a factor $\omega_p^2/\omega_0^2$ relative to the latter. So we may neglect the second term and simplify the formula as 
\begin{eqnarray}
\frac{\ddot{x}_{_M}}{c}&=&\frac{1-(1/2)\omega_p^2/\omega_0^2}{[1-(3\pi/2) c\omega_p'/\omega_p^2]^3} \times \\  \nonumber
&&\Big\{-\frac{\omega_p^2}{\omega_0^2}\frac{\omega_p'}{\omega_p}
+\frac{3\pi c}{2}\Big[\frac{\omega_p''}{\omega_p^2}-2\frac{\omega_p'^2}{\omega_p^3}\Big]\Big\}.
\end{eqnarray}

Our desire is to achieve as high an acceleration as possible. To accomplish that, one should design the system in such a way that the characteristic length $D$ and the denominator are minimized. 

\section{\label{sec:analogBH}Application to Analog Black Holes}

For the purpose of investigating Hawking evaporation and the information loss paradox, it is desirable to maximize the mirror acceleration. The laser frequency $\omega_0$ is largely determined by the laser technology and so there is not a large room for tuning. About the plasma frequency $\omega_p$, the range is much larger. Essentially all the plasma based wakefield accelerator \cite{Tajima:1979,Chen:1985} experiments have been invoking gaseous plasmas. In the Chen-Mourou proposal \cite{Chen:2017}, extremely high density plasmas induced by solid targets through nano-fabrication technology was proposed. 
To be sure, there are pros and cons between these two types of plasmas. One obvious drawback of the nano-target is the inevitable plasma-induced background events that would compete with the Hawking signals. 

In addition to the plasma frequency, one other important parameter is the characteristic length of the plasma density gradient, $D$, associated with $\omega_p'$ and $\omega_p''$, which plays an essential role in contributing to the value of mirror acceleration $\ddot{x}_{_M}$. As we will see below, this turns out to be the most sensitive parameter that controls the Hawking temperature. Furthermore, the signs of both $\omega_p'$ and $\omega_p''$ are evidently critical to the final value of  $\ddot{x}_{_M}$. As discussed in the previous section, we shall focus on the down ramp case.

\vskip 0.2 in

\noindent{1. \it{Exponential Profile}}

A simple but well motivated plasma density profile is the one that corresponds to the exponential trajectory investigated by Davies and Fulling \cite{Fulling-Davies:1976,Fulling-Davies:1977}, which is of special geometrical interest because it corresponds to a well-defined horizon \cite{Birrell:1983}. Inspired by that, we consider the following plasma density variation along the direction of the laser propagation:
\begin{eqnarray}
n_p(x)=n_{p0}(a+be^{-x/D})^2, \quad\quad 0\leq x  \leq x_b,
\end{eqnarray}
where $n_{p0}$ is the plasma density at $x=0$, $D$ is a characteristic length of density variation. Accordingly, the plasma frequency varies as 
\begin{eqnarray}
\omega_p(x)=\omega_{p0}(a+be^{-x/D}), \quad\quad 0\leq x  \leq x_b,
\end{eqnarray}
where $\omega_{p0}=c\sqrt{4\pi r_e n_{p0}}$. In our conception, the time derivatives of the plasma frequency are induced through the spacial variation of the plasma density via the relation $\omega_p(x)=c\sqrt{4\pi r_e n_p(x)}$. Thus
\begin{eqnarray}
\omega'_p(x)&=&-\frac{b}{D}e^{-x/D}\omega_{p0}(x), \\ 
\omega''_p(x)&=&\frac{b}{D^2}e^{-x/D}\omega_{p0}(x).
\end{eqnarray}
Inserting these into Eq.(20) and Eq.(25), we then have, for the exponential distribution, 
\begin{eqnarray}
\frac{\dot{x}_{_M}}{c}=
\frac{1-(\omega_{p0}^2/2\omega_0^2)(a+be^{-x/D})^2}
{1+(3b/4)(\lambda_{p0}/D)e^{-x/D}/(a+be^{-x/D})^2}.
 \end{eqnarray}
and
\begin{eqnarray}
\frac{\ddot{x}_{_M}}{c^2}&=&\frac{1-(\omega_{p0}^2/2\omega_0^2)(a+be^{-x/D})^2}
{[1+(3b/4)(\lambda_{p0}/D)e^{-x/D}/(a+be^{-x/D})^2]^3}  \\     \nonumber
&&\times \frac{\lambda_{p0}}{D}\frac{be^{-x/D}}{(a+be^{-x/D})} \Big\{\frac{\omega_p^2}{\omega_0^2}\frac{1}{\lambda_{p0}} \\  \nonumber
&&+ \frac{3}{4D}\Big[\frac{1}{a+be^{-x/D}}-\frac{2be^{-x/D}}{(a+be^{-x/D})^2}\Big]\Big\}. \nonumber
\end{eqnarray}

\noindent{2. \it{Gaussian Profile}}

A common practice in plasma wakefield acceleration experimentation is to invoke gas jets as the plasma target, where the intersecting laser would instantly ionize the neutral gas into plasma. The density profile of a gas jet is usually in Gaussian distribution. Motivated by the practical experimentation consideration, here we investigate the case of  a half-Gaussian plus a constant profile, defined as 
\begin{eqnarray}
n_p(x)=n_{p0}(a+be^{-x^2/2D^2})^2, \quad\quad 0\leq x  \leq x_b,
\end{eqnarray}
where $n_{p0}$ is the plasma density at $x=0$, $D$ is a characteristic length of density variation and $x_b > D$ is the location of the plasma front and back boundary. Accordingly, the plasma frequency varies as 
\begin{eqnarray}
\omega_p(x)=\omega_{p0}(a+be^{-x^2/4D^2}), \quad\quad 0\leq x  \leq x_b,
\end{eqnarray}
where $\omega_{p0}=c\sqrt{4\pi r_e n_{p0}}$. In our conception, the time derivatives of the plasma frequency are induced through the spacial variation of the plasma density via the relation $\omega_p(x)=c\sqrt{4\pi r_e n_p(x)}$. Thus
\begin{eqnarray}
\omega'_p(x)&=&-b\frac{x}{D^2}e^{-x^2/2D^2}\omega_{p0}(x), \\ 
\omega''_p(x)&=&\frac{b}{D^2}\Big(-1+\frac{x^2}{D^2}\Big)e^{-x^2/2D^2}\omega_{p0}(x).
\end{eqnarray}
Inserting these into Eq.(20) and Eq.(25), we then have, for Gaussian distribution, 
\begin{eqnarray}
\frac{\dot{x}_{_M}}{c}=
\frac{1-\omega_p^2/2\omega_0^2}
{1+(3b/2)(\lambda_{p0} x/D^2)e^{-x^2/2D^2}},
 \end{eqnarray}
and
\begin{eqnarray}
\frac{\ddot{x}_{_M}}{c^2}&=&\frac{1-\omega_p^2/2\omega_0^2}
{[1+(3b\lambda_{p0} x/4D^2)e^{-x^2/2D^2}/(a+be^{-x^2/2D^2})^{2}]^3}  \\     \nonumber
&&\times \frac{b\lambda_{p0}}{2D^2}\frac{e^{-x^2/2D^2}}{(a+be^{-x^2/2D^2})^2} \Big\{\frac{\omega_p^2}{\omega_0^2}\frac{x}{\lambda_{p0}} \\  \nonumber
&&+ \frac{3}{4}\Big[-1+\frac{x^2}{D^2}-\frac{2bx^2}{D^2}\frac{e^{-x^2/2D^2}}{a+be^{-x^2/2D^2}}\Big]\Big\}. \nonumber
\end{eqnarray}


\section{\label{sec:max}Analog Hawking Temperature}

There exists a wealth of literature on the mode function of the quantum field, their reflections from a flying mirror, and the analog ``Hawking temperature" of such a flying mirror as an analog black hole \cite{Birrell:1983}. In general it depends on the actual mirror trajectory. Starting with Eq.(22), we obtain
\begin{eqnarray}
\int_{t_0}^{t}\bar{c}_0 dt=
\int_{x_0}^{x}\Big[1+\frac{3b\lambda_{p0}}{4D}\frac{e^{-x/D}}{(a+be^{-x/D})^2}\Big]dx,
\end{eqnarray} 
where $\bar{c}_0=\eta_0 c=(1-\omega_{p0}^2/2\omega_0^2)c$ is the speed of light in the plasma medium. In the $t\to \infty$ (also $x\to \infty$) limit, we have 
\begin{eqnarray}
x_{_M}(t)&=&\bar{c}_0t-\frac{3b}{4a^2}\lambda_{p0}\big[e^{-\bar{c}_0t/D}-e^{-\bar{c}_0t_{0}/D}\big]  \\ 
&\equiv&\bar{c}_0t-Ae^{-\bar{c}_0t/D}+B. 
 \end{eqnarray}
Transcribing the $(x,t)$ coordinates to the $(u,v)$ coordinates, where $u=\bar{c}_0t-x$ and $v=\bar{c}_0t+x$, we see that only null rays with $v<B$, where $B$ is the constant term in the mirror trajectory (see below), can be reflected. All rays with $v>B$. will pass undisturbed. The ray $v=B$ therefore acts as an effective horizon \cite{Birrell:1983}. Following the standard recipe \cite{Birrell:1983}, we obtain the Wightman function as 
\begin{eqnarray}
D^+(u,v;u',v')&=&-\frac{1}{4\pi}\ln\big[2\eta_0e^{2B\eta_0/D}e^{\eta_0(t+t')/2}\big] \cr
&\times& \sinh(\eta_0\Delta t/2D),
\end{eqnarray}
where $\Delta t=t-t'=\Delta u/2$ in the $t\to \infty$ limit. The constant factors in the argument of the log  function in the above equation do not contribute to the nontrivial part of the physics. We then have, in the asymptotic limit of $t, t' \to \infty$, 
\begin{eqnarray}
D^+(u,v;u',v')=-\frac{1}{4\pi}\ln\big[\sinh(\eta_0\Delta t/2D)\big].
\end{eqnarray} 
This leads to the response function (of the particle detector) per unit time with the form
\begin{eqnarray}
\mathcal{F}(E)/{\rm unit \ time}=\frac{1}{E}\frac{1}{(e^{E/k_{_B}T_{H}}-1)},
\end{eqnarray}
where the analog Hawking temperature of the mirror measured by a stationary particle detector is
\begin{eqnarray}
k_{_B}T_{_H}=\frac{\hbar c}{2\pi}\frac{\eta_0}{D}.
\end{eqnarray}
where $k_{_B}$ is the Boltzmann constant.

Consider a flying plasma mirror experiment where the driving laser has the frequency $\omega_0=3.5 \times 10^{15} {\rm sec}^{-1}$. For the plasma target, we set $a=b=1$, and $n_p(x)=n_{p0}(1+e^{-x/D})$, $n_{p}(x=0)=1.0 \times 10^{17}{\rm cm}^{-3}=4n_{p0}$. The corresponding plasma frequency is $\omega_{p0}=0.9 \times 10^{13}{\rm sec}^{-1}$ and the plasma wavelength $\lambda_{p0}=100 \mu{\rm m}$. Next we design the plasma target density profile in such a way that $D=0.5\mu{\rm m}$. Then from the above equation we find
\begin{eqnarray}
k_{_B}T_{_H}\sim 6.6 \times 10^{-2} {\rm eV},
\end{eqnarray}
which corresponds to a characteristic Hawking radiation frequency $\omega_{_H} \sim 1.3\times 10^{13}{\rm sec}^{-1} > \omega_{p0}$. So the Hawking radiation can propagate through the plasma for detection.

\section{\label{sec:dis}Conclusion}

One key element in the concept of flying plasma mirrors as analog black holes is the dynamics of the mirror acceleration induced by the plasma target density gradient. In this paper we fill in the gap of such dynamical details left by the original proposal of Chen and Mourou \cite{Chen:2017}. Different physical effects that cause the plasma mirror to slow down or speed up are explicitly analyzed. Based on such microscopic examination, the mirror speed, trajectory, and acceleration are derived. While the numerical example provided invoked a gaseous plasma, our formulas are generic that can be applied to solid plasmas as well.

As mentioned in the Introduction, from the experimentation's point of view, one of the challenges of the proposed two-stage analog black hole experiment is the excessive backgrounds generated by the nano-thin-film target in the second stage. Another is the preparation of a high-intensity x-ray pulse to induce the flying plasma mirror inside the nano-target. In this work we have shown that a single-stage, gaseous plasma target is viable. This would help to bypass the challenges of the intense x-ray production and the nano-target fabrication. It is hoped that the results in this paper would provide a guidance to the design and operation of experiments using flying plasma mirrors as analog black holes.

\section*{Acknowledgment}

The author appreciate helpful discussions with Timur Esirkepov of Kansai Photon Science Institute and Chih-En Chou, Patrick Yuan Fang, Kuan-Nan Lin, Yung-Kun Liu, and Prof. Stathes Paganis of National Taiwan University. This work is supported by ROC Ministry of Science and Technology (MOST), National Center for Theoretical Sciences (NCTS), and Leung Center for Cosmology and Particle Astrophysics (LeCosPA) of National Taiwan University. He is in addition supported by U.S. Department of Energy under Contract No. DE-AC03-76SF00515.


\end{document}